\input{aipcheck}

\documentclass[
    ,final            
  ]
  {aipproc}

\layoutstyle{8x11single}

\usepackage{amssymb}
\def\Journal#1#2#3#4{{#1} {\bf #2}, #3 (#4)}

\def\PLB{{\em Phys. Lett.}  B}

\def\PRL{\em Phys. Rev. Lett.}

\def\PRD{{\em Phys. Rev.} D}

\begin{document}

\title{Exotic Searches at the Tevatron}


\author{Gustaaf Brooijmans \\ {\it on behalf of the CDF and D\O\ Collaborations}}{
  address={Columbia University, New York, NY}
}



\begin{abstract}
Recent results on searches for new physics at Run II of the Tevatron are reported.
The searches cover many different final states and previous hints of signals, but 
all analyses have at this point led to negative results.
\end{abstract}

\maketitle


\section{Introduction}
In the standard model of particle physics, the $W_L W_L$ scattering amplitude violates 
the unitarity bound at a center of mass energy c.o.m. $\approx 1.7$ TeV~\cite{quigg}.
One solution to this problem is offered by the Higgs mechanism~\cite{higgs} through the 
introduction of a massive scalar particle with non-zero vacuum expectation value.
Loop corrections to this ``Higgs boson's'' mass involving other massive particles are 
quadratically divergent, implying that the physical mass of the Higgs boson corresponds to a
scale at which new physics is present.  To successfully address the $W_L W_L$ scattering 
amplitude problem the Higgs boson mass is constrained to $m_H \lessapprox 1$ TeV/$c^2$, and if 
fermions acquire their masses through coupling to the Higgs boson, then 
$m_H \lessapprox 200$ GeV/$c^2$ \cite{lepewwg}.  If the Higgs boson, whether elementary or 
composite, doesn't exist, some other form of new physics must be present at the TeV scale
to prevent the $W_L W_L$ scattering amplitude from violating the unitarity bound.  
Among the models of new physics, supersymmetric ones are certainly, and justifiably, the most 
popular.  These address the divergences in the corrections to the Higgs mass in a very 
natural way, improve the unification of couplings at a high scale, provide a good
dark matter candidate and lead to radiative electroweak symmetry breaking.  
Unfortunately, they do not tell us anything about the 
origin of the number of generations of fermions, their mass spectrum, charges, etc.  In 
this talk, recent results from the CDF and D\O\ collaborations on searches for manifestations 
of non-supersymmetric new physics are presented.

\section{New Gauge Bosons}

New heavy gauge bosons are predicted in many extensions of the standard model:
\begin{itemize}
\item In so-called Little Higgs models, the quadratically divergent radiative corrections
 to the Higgs mass are canceled ``individually'', leading to the appearance of partners
 of the $W$ and $Z$ bosons at the TeV scale.
\item In Grand Unified Theories (GUTs) and Left-Right Symmetric Models (LRSMs) heavy partners
 of the electroweak bosons generally appear, although their masses are less well constrained.
\item In extra dimensional models where the gauge bosons are allowed to propagate in (some of) 
 the additional space dimensions Kaluza-Klein excitations will be observable.  To address the 
 hierarchy problem, these should typically have masses $m \lessapprox 10$ TeV/$c^2$.
\item If the Higgs field is a triplet rather than a doublet, doubly charged Higgs bosons
 exist. 
\end{itemize}
It should also be noted that graviton resonances which appear in Randall-Sundrum type 
models~\cite{rs} lead to similar signatures.

\begin{figure}
  \includegraphics[height=.3\textheight]{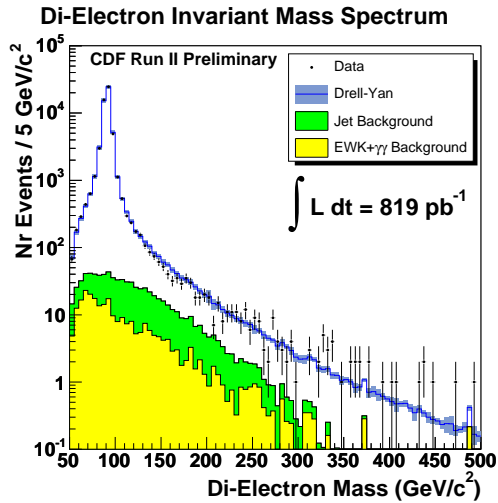}
  \caption{\label{fig:cdf-diem} Dielectron invariant mass spectrum in the CDF search 
for electron-positron resonances.}
\end{figure}

CDF recently released a new result in the search for dielectron resonances using almost
900 $pb^{-1}$ of data.  The invariant
mass spectrum of dielectron events used in this search is shown in Figure~\ref{fig:cdf-diem}.
No significant excess is seen at any mass value leading to a limit on the mass of 
a sequential, standard model-like $Z'$ of $m(Z') > 850$ GeV$/c^2$ at 95\% C.L.  

CDF 
also searches for pair production of doubly charged Higgs bosons in the lepton 
flavor violating modes $H^{++} \rightarrow e^+ \tau^+$ and  
$H^{++} \rightarrow \mu^+ \tau^+$ (and charge conjugates). At least three leptons
are required in the final states so that the backgrounds are very small, and no events are
observed in approximately 350 $pb^{-1}$ of data.  Interpreted in a left-right symmetric 
model, the corresponding mass limit on doubly 
charged Higgs bosons coupling to left-handed particles is $m(H^{++}) > 114 (112)$ GeV$/c^2$
at 95\% C.L. in the $e\tau (\mu\tau)$ channel.

In the search for singly charged gauge bosons, CDF looked for a $W'$ decaying to an 
electron and a neutrino in 205 $pb^{-1}$ and derived a limit $m(W') > 788$ GeV$/c^2$ 
from a study of the transverse mass spectrum.

\section{Leptoquarks}

Leptoquarks are a natural consequence of the unification of quarks and leptons into 
a single multiplet, and as such are expected to be gauge bosons as well.  While their
masses can logically be expected to be of the order of the unification scale, in 
some models they can be relatively light.  Experimentally, it is customary to 
consider one leptoquark per generation.  These are assumed to be very short-lived and
decay to a quark and a lepton.  The branching ratio to a charged lepton and a quark
is then denoted as $\beta$.  

At hadron colliders, leptoquarks can be pair-produced through the strong 
interaction or singly produced. In the latter case the production cross-section depends 
on the (unknown) quark-lepton coupling, which is generally taken to be of the same 
order of magnitude as the fine structure constant.

\begin{figure}
  \includegraphics[height=.3\textheight]{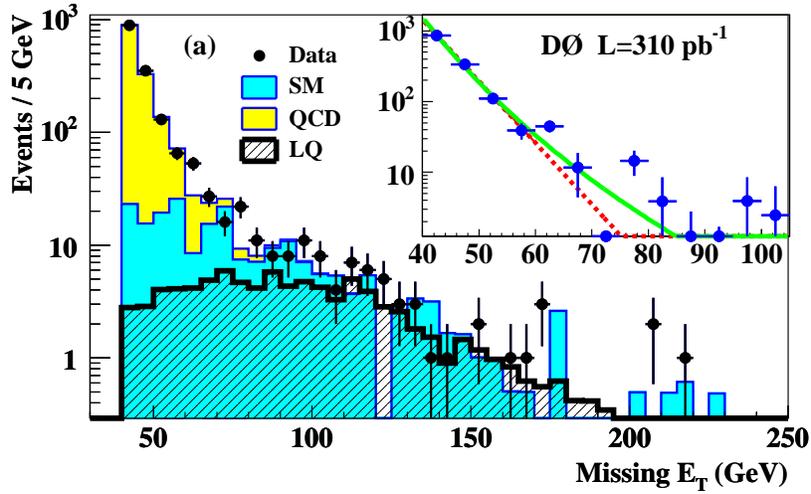}
  \caption{\label{fig:d0-met} Missing transverse energy distribution in the D\O\
search for $\beta = 0$ leptoquarks before final cuts.  The inset shows the two 
different fitting functions used to evaluate the background from mismeasured 
QCD dijet events.}
\end{figure}

There are two recent results from leptoquark searches at the Tevatron.  D\O\ has 
searched for leptoquarks decaying to a quark and a neutrino ($\beta = 0$) in 
the jets plus missing tranverse energy in 370 $pb^{-1}$ of data.  Experimentally 
this is a difficult 
analysis which suffers from substantial QCD dijets background due to mismeasured
jets.  To mitigate this, D\O\ requires exactly two acoplanar jets.  The ensuing
missing transverse energy distribution, before final analysis cuts, is shown in 
Figure~\ref{fig:d0-met}.
The background from QCD dijets, dominant at low missing transverse energy, 
is extrapolated to higher values using two different fitting functions as 
shown in the inset.  The dominant non-QCD standard model background is 
$Z$ boson plus jets production with the $Z$ decaying to a pair of neutrinos.
No excess is observed so D\O\ sets a limit at $m(LQ) > 136$ GeV$/c^2$ at 95\% C.L.

CDF has released results on a search for vector leptoquarks of the third 
generation in the $LQ_3 \rightarrow b\tau$ decay channel.  The signature 
consists of a dijet plus ditau final state, in which one tau is required to decay
leptonically and the other hadronically.  The main discriminating variables 
are the number of jets and an $H_T$-type variable which is a scalar sum of the 
transverse energies of all jets, leptons and the event's missing transverse 
energy.  This allows CDF to set a limit at $m(LQ_3) > 294$ GeV$/c^2$ assuming
$\beta = 1$.  Note that this limit is much higher than the typical limits
on leptoquark masses at the Tevatron due to the model choice of vector leptoquarks,
which have a much larger production cross-section than the scalar leptoquarks
which are usually chosen.

\section{Large Extra Dimensions}

In the original large extra dimensions model of Arkani-Hamed, Dimopoulos and 
Dvali~\cite{add} in which only gravitons propagate in the bulk but all 
standard model fields are confined to a 3-brane, a tower of Kaluza-Klein
excitations of the graviton emerges.  The graviton states are too close 
in mass to be distinguished individually, and the coupling remains small, 
but the number of accessible states is very large.  It is therefore possible 
to produce gravitons which immediately disappear into bulk space, leading 
to an excess of events with a high transverse energy jet and large missing
transverse energy, the so-called monojet signature.  The dominant standard 
model backgrounds are the production $Z$ or $W$ bosons plus jets, with the 
$Z$ decaying to a pair of neutrinos or the lepton from $W$ decay escaping 
detection.  Using 368 $pb^{-1}$ of data, CDF sets limits on the fundamental
Planck scale between $M_D > 1.16$ TeV$/c^2$ and $M_D > 0.83$ TeV$/c^2$ for a number
of extra dimensions ranging from 2 to 6.

\section{Other Resonances}

Both CDF and D\O\ have new results on the search for resonances decaying to $Z + X$.
CDF has two analyses, with the first using $H_T$ as its main discriminating 
variable.  After selecting events with leptonic $Z$ decays, a control region with 
$H_T < 200$ GeV is used to establish a good
understanding of the data, and the signal region is chosen to have $H_T > 400$ GeV, 
including at least two jets, each with transverse energy $E_T > 50$ GeV.  Using 
305 $pb^{-1}$ of data, no events are observed in the signal region.  A limit is set 
on the production cross-section of $m(D) = 300$ GeV$/c^2$ right-handed down-type 
quarks as proposed in reference~\cite{bj} at $\sigma < 1.3$ pb at 90\% C.L.

\begin{figure}
  \includegraphics[height=.3\textheight]{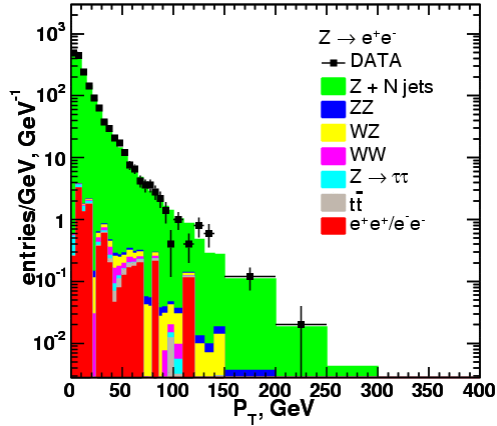}
  \caption{\label{fig:cdf-zpt} Transverse momentum distribution of $Z$ bosons decaying
to electron-positron pairs measured by CDF in 305 $pb^{-1}$ of data.}
\end{figure}

In a second analysis, CDF studies the transverse momentum distribution of $Z$ bosons.
This has the advantage of being insensitive to the nature of the other decay products,
but is of course more difficult and potentially less sensitive than a direct 
resonance search.  Selecting events with a dielectron invariant mass compatible with 
the mass of the $Z$ boson, $66$ GeV$/c^2 < M(ee) < 116$ GeV$/c^2$, CDF measures the $Z$
transverse momentum distribution shown in Figure~\ref{fig:cdf-zpt}.
From this they determine an upper limit on the anomalous production of $Z$ bosons as a
function of transverse momentum using 305 $pb^{-1}$ of data.  The 95\% C.L. limit ranges 
from about 
1 pb for $p_T(Z) = 20$ GeV$/c$ to approximately 2 fb for $p_T(Z) = 200$ GeV$/c$.

\begin{figure}
  \includegraphics[height=.3\textheight]{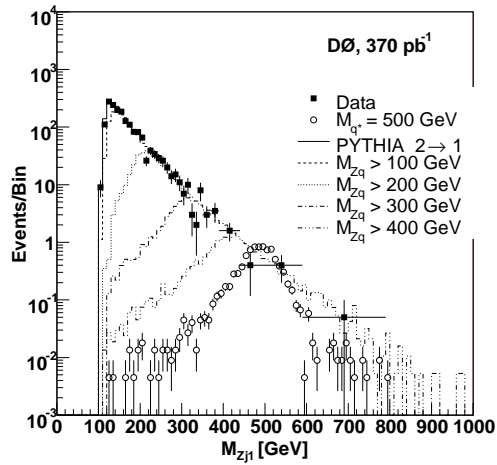}
  \caption{\label{fig:d0-zjm} Invariant mass of the $Z$ boson and leading jet as 
measured by D\O\ in 380 $pb^{-1}$ of data. The curves with different $M_{Zq}$ thresholds
correspond to different generation thresholds for 2 $\rightarrow$ 2 processes in Pythia,
and the open circles represent the signal due to an excited quark of mass 
$m_{q^*} = 500$ GeV$/c^2$ and narrow width.}
\end{figure}

D\O\ explicitely searches for a $Z$ boson plus jet resonance by combining the 
$Z$ boson plus jet mass spectrum and the $Z$ boson transverse momentum distribution 
as discriminating variables.  The invariant mass distribution measured in 380 $pb^{-1}$
of data is shown in Figure~\ref{fig:d0-zjm}.
No excess is observed and a limit is set on the mass of an excited quark as proposed
in Reference~\cite{qs} at $m_{q^*} > 520$ GeV$/c^2$ at 95\% C.L.

\begin{figure}
  \includegraphics[height=.3\textheight]{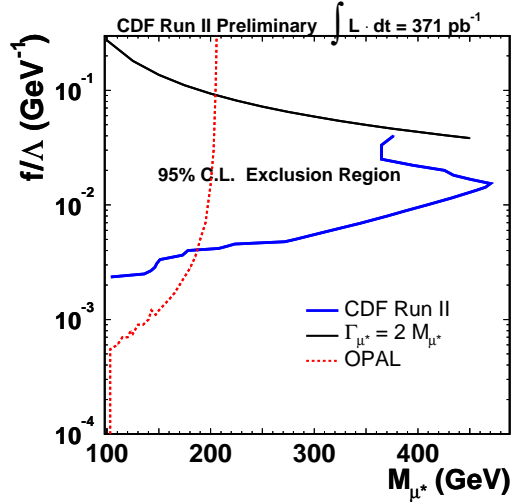}
  \caption{\label{fig:cdf-muslim} Exclusion region from the CDF search for excited 
muons in a gauge mediated model with coupling $f/\Lambda$ for comparison with 
LEP results.}
\end{figure}

In the search for excited muons both CDF and D\O\ have 
results as well.  Both experiments search for associated production of a muon and 
an excited muon, with the latter decaying to a muon and a photon.  The production 
is approximated as a contact interaction, while the decay is assumed to proceed
either through a gauge interaction (CDF) or a combination of a gauge and a contact
interaction, with the relative fraction of the two depending on the mass of the 
excited muon and the compositeness scale.  Both experiments obtain very similar 
results using 371 (CDF) and 380 (D\O) $pb^{-1}$ of data.  To make comparison 
with LEP results easier, CDF also reinterprets the result in a  
gauge mediated model with Drell-Yan-like production of the $\mu \mu^*$ pair with 
coupling $f/\Lambda$.  This result is shown in Figure~\ref{fig:cdf-muslim}.

\section{Events with Leptons and Photons}

In Run I, CDF reported an excess of events with a photon, a lepton and large 
missing transverse energy compared to standard model expectations~\cite{cdflg}.  
This excess, corresponding to a 2.7 sigma effect, was deemed ``an interesting 
result'', but not ``a compelling observation of new physics.''  CDF has repeated
this analysis in Run II using 305 $pb^{-1}$ of data~\cite{cdflg2}.  The data is 
now compatible with standard model expectations in all channels, suggesting that
the excess observed in Run I was due to a statistical fluctuation.

In a similar spirit, CDF searches for events with two photons and an electron or
muon (700 $pb^{-1}$), or three photons (1 $fb^{-1}$).  This is primarily a 
counting experiment, and again, unfortunately, no excess is observed over expectations.

\section{Displaced Dimuons}

\begin{figure}
  \includegraphics[height=.3\textheight]{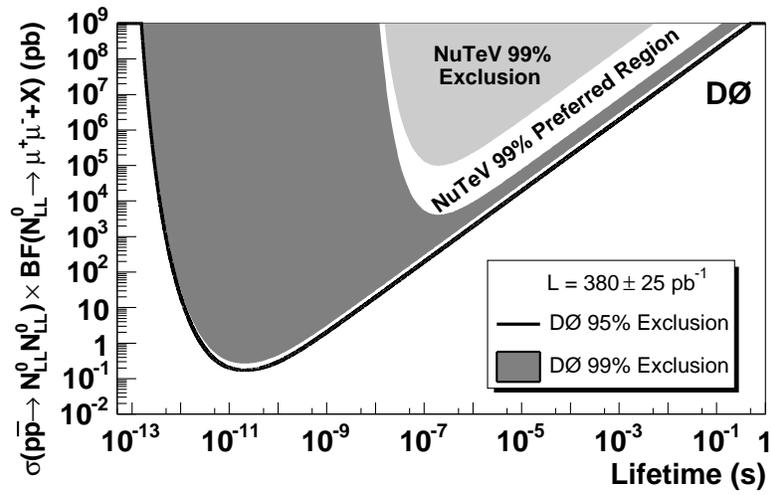}
  \caption{\label{fig:d0-nutev} Excluded area in the search for a long-lived particle 
decaying to muon pairs plus anything in terms of the particle's production cross-section
times branching ratio, and lifetime.}
\end{figure}

In 2000, The NuTeV Collaboration reported~\cite{nutev} on a search for heavy 
neutral leptons decaying to $\mu \mu \nu$ and a few other final states.  In the 
dimuon channel, they observed three events while only 0.07 were expected.  Because
of the asymmetry in the muon momenta, and the absence of signal in other channels,
NuTeV argued that the signal was unlikely to be due to neutral heavy leptons.
To further investigate this, D\O\ has searched for pairs of muons originating 
from a common vertex located between 5 and 20 $cm$ from the beamline.  This allows
a good calibration of efficiencies using $K_S$ mesons.  No signal is seen in 
380 $pb^{-1}$ of data, and 
to compare results the momentum of the hypothetical new particles in the neutrino 
beam was converted to the Tevatron center of mass frame.  The result, which is weakly
dependent on assumptions made in regard to the new particle's decay process, is 
illustrated in Figure~\ref{fig:d0-nutev} in terms of the particle's production 
cross-section times branching ratio and lifetime.

\section{Conclusions}

As the Tevatron continues its good performance, CDF and D\O\ collect large amounts of 
data and search for new physics in many channels.  Only recent results have been reported 
here, and many other analyses are in progress.  There is {\it still} nothing interesting,
and previous excesses have been shown to be due to fluctuations.  So far, new physics 
has remained hidden, but a substantial amount of data is yet to be recorded at the Tevatron
and the LHC is scheduled to start operations soon.  Positive results should be obtained in 
the next few years.

\end{document}